\documentclass[dvips]{arxstspdf}
\usepackage{flushend}
\usepackage{stfloats}
\usepackage{graphics}
\volume{24}
\issue{3}
\pubyear{2009}
\firstpage{328}
\lastpage{342}
\doi{10.1214/09-STS305}

\makeatletter
  \let\sv@tabnotetext\tabnotetext
  \let\sv@tabnotemark@fmt\tabnotemark@fmt
   \long\def\legend#1{{\let\tabnote@indent\leavevmode\sv@tabnotetext[]{}{#1}}}
\makeatother

\begin{document}
\begin{frontmatter}

\title{Interval Estimation for Messy Observational Data}
\runtitle{Interval Estimation}

\begin{aug}
\author[a]{\fnms{Paul} \snm{Gustafson}\ead[label=e1]{gustaf@stat.ubc.ca}\corref{}} \and
\author[b]{\fnms{Sander} \snm{Greenland}\ead[label=e2]{lesdomes@ucla.edu}}
\runauthor{P. Gustafson and S. Greenland}

\affiliation{University of British Columbia and University of California}

\address[a]{Paul Gustafson is Professor, Department of Statistics,
University of British Columbia,
Vancouver, BC V6T 1Z2, Canada
\printead{e1}.}
\address[b]{Sander Greenland is Professor,
Departments of Epidemiology and Statistics
University of California,
Los Angeles, CA 90095-1772, USA
\printead{e2}.}

\end{aug}

\begin{abstract}
We review some aspects of Bayesian and frequentist interval
estimation, focusing first on their relative strengths and weaknesses
when used in ``clean'' or ``textbook'' contexts. We then turn attention to
observational-data situations which are ``messy,'' where modeling that
acknowledges the limitations of study design and data collection leads
to nonidentifiability. We argue, via a series of examples, that Bayesian
interval estimation is an attractive way to proceed in this context even
for frequentists, because it can be supplied with a diagnostic in the
form of a calibration-sensitivity simulation analysis. We illustrate the
basis for this approach in a series of theoretical considerations,
simulations and an application to a study of silica exposure and lung
cancer.
\end{abstract}

\begin{keyword}
\kwd{Bayesian analysis}
\kwd{bias}
\kwd{confounding}
\kwd{epidemiology}
\kwd{hierarchical prior}
\kwd{identifiability}
\kwd{interval coverage}
\kwd{observational studies}.
\end{keyword}

\end{frontmatter}

\section{Introduction}

The~conventional approach to observational-data analysis is to apply
statistical methods that assume a designed experiment or survey has been
conducted. In other words, they assume that all unmodeled sources of
variation are randomized under the design. In most settings, deviations
of the reality from this ideal are dealt with informally in
post-analysis discussion of study problems. Unfortunately, such informal
discussion seldom appreciates the potential size and interaction of
sources of bias and, as a consequence, the conventional approach
encourages far too much certainty in inference (Eddy, Hasselblad and
Schachter, \citeyear{EdHaSc1992}; Greenland, \citeyear{Gr2005}, \citeyear{Gr2009}; Greenland and Lash, \citeyear{GrLa2008};
Molitor et al., \citeyear{MoJaBeRi2009}; Turner et al., \citeyear{TuSpSmTh2009}).

The~entrenchment of the conventional approach derives in part from the
fact that realistic models for observational studies are not identified
by the data, a~fact which renders conventional methods and software
useless (except perhaps as part of a larger fitting cycle). The~most
commonly proposed mode of addressing this problem is sensitivity
analysis, which, however, leads to problems of dimensionality and
summarization. The~latter problems have in turn been addressed by
Bayesian and related informal simulation methods for examining
nonidentified models (which are often dealt with under the topic of
nonignorability). These methods include hierarchical (multilevel)
modeling of biases (Greenland, \citeyear{Gr2003}, \citeyear{Gr2005}), which is intertwined with
the theme of the present paper.

We start in Section \ref{sec2} by reviewing some notions of interval estimator
performance, with emphasis on coverage averaged over \textit{different}
parameter values. Section \ref{sec3} then extends this discussion to include
intervals arising from \textit{hierarchical} Bayesian analysis when data
from multiple studies are at hand. These two sections reframe existing
theory and results in a manner suited for our present needs. We
emphasize a well-known tradeoff: To the extent the selected prior
distribution is biased relative to reality, the coverage of a Bayesian
posterior interval will be off, but perhaps not by much; and in return
the intervals can deliver substantial gains in precision and reduced
false-discovery rates compared to frequentist confidence intervals. In
addition, hierarchical priors provide a means to reduce prior
misspecification as studies unfold.

In Section \ref{sec4} we turn to the more novel aspect of our work, by studying
the case which we believe better captures observational-study reality,
in which \mbox{priors} are essential for identification. Here the usual order
of robustness of frequentist vs. Bayesian procedures reverses:
Confidence intervals become only extreme posterior intervals, obtained
under degenerate priors, with coverage that rapidly deteriorates as
reality moves away from these point priors. In contrast, the general
Bayesian framework with proper priors offers some protection against
catastrophic undercoverage, with good coverage guaranteed under a
spectrum of conditions specified by the investigator and transparent to
the consumer. Section \ref{sec5} summarizes the lessons we take away from our
observations and makes a recommendation concerning the practical
assessment of interval estimator performance. We conclude that Bayesian
interval estimation is an attractive way to proceed even for
frequentists, because its relevant calibration properties can be checked
in each application via simulation analysis. We close with an
illustration of our proposed practical approach in an application to a
study of silica exposure and lung cancer in which an unmeasured
confounder (smoking) renders the target parameter nonidentified.

\section{The~Well-Calibrated Lab}\label{sec2}

Let $\theta$ denote the parameter vector, and $D$ the observable data,
for a study that is to be carried out. Assume for now that the
distribution of ($D|\theta) ($i.e., ``the model'') is known correctly.
Say that $\phi =g(\theta)$ is the scalar parameter of interest, and
that $I(D)$ is an interval estimator for this target. We define the
\textit{labwise coverage} (LWC) of $I$ with respect to a
\textit{parameter-generating distribution (PGD) P} as
\begin{equation}\label{for1}
C(I,P) = \operatorname{Pr}\{\phi \in{I(D)}\}.
\end{equation}
Here the probability is taken with respect to the distribution of
$(\theta, D)$ jointly, with $\theta \sim P$ and $(D | \theta)$ following
the model distribution.

Interval coverage with respect to a joint distribution on parameters and
data, as in (\ref{for1}), has been considered by many authors, but not with a
consistent terminology. While it might be temping to refer to (\ref{for1}) as
``Bayesian'' coverage, we find this confusing since (\ref{for1}) can be
evaluated for Bayesian or non-Bayesian interval estimators. We choose to
call it labwise coverage since $C( I,P)$ is the proportion of
right answers reported by a lab or research team applying estimator $I$
in a long series of studies of \textit{different} phenomena (different
exposure-disease relationships, say) within a research domain. The~role
of the PGD $P$ is then to describe the corresponding across-phenomena
variation in the underlying parameter values. Interest in labwise
coverage might be very direct in some contexts, in that estimator
operating characteristics in a long sequence of actual studies really
are the primary consideration. Or interest may be more oblique, in that
performance on the ``next'' study is of interest, and this performance
is being measured conceptually by regarding the next study as a random
draw from the population of ``potential'' or ``future'' studies.

If $I$ is a frequentist confidence interval (abbreviated FCI), then it
will attain nominal coverage exactly for any PGD. That is, if $\operatorname{Pr}\{\phi
\in {I(D)} | \theta\} = 1 - \alpha$ for every value of
$\theta$, then $C(I,P)=1 - \alpha$ for any $P$. Thus, correct
coverage for a hypothetical sequence of studies with the same parameter
values implies correct coverage in the more realistic setting of
repeatedly applying a procedure in a sequence of differing real
problems. While this fact is often viewed as a robustness property of an
FCI, Bayarri and Berger (\citeyear{BaBe2004}), citing Neyman (\citeyear{Ne1977}), emphasize that it
is the labwise coverage that is relevant for practice. Put another way,
if a lab is well calibrated in the LWC sense of producing 95\% intervals
that capture the true parameter for 95\% of studies, and the cost of
failing to capture is the same across studies (as might be the case in
some genome studies or screening projects), there is little obvious
benefit if the intervals happen to also have correct frequentist
coverage.

\subsection{Bayesian Intervals under PGDs}

For a given choice of prior distribution $\Pi$ on the parameter vector
$\theta$, a $1 -\alpha$ Bayesian posterior credible interval (BPCI)
for the target parameter $\phi$ would be any interval having Bayesian
probability $1 - \alpha$ of containing $\phi$ given the observed data
$D$. The~most common choices of BPCI are the \textit{equal-tailed} BPCI
(i.e, the \mbox{interval} formed by the $\alpha /2$ and $1-\alpha /2$ posterior
quantiles of the target parameter), and the
\textit{highest-posterior-density} (HPD) BPCI. Though HPD intervals are
optimally short, we consider only equal-tailed intervals here, given
their simple interpretation and widespread use.

If the prior $\Pi$ and the PGD $P$ coincide, then a BPCI is guaranteed
to have correct labwise coverage. This strikes us as a fundamental
property of BPCIs, though it is surprisingly unemphasized in most
introductions to Bayesian techniques. Henceforth, we refer to a BPCI
arising from a prior distribution set equal to the PGD as an
\textit{omniscient} or ``oracular'' BPCI (abbreviated OBPCI), in the
sense that the investigator is omniscient in knowing the actual PGD
giving rise to future studies. It is indeed a fanciful assumption to
think that the PGD would be known exactly, so throughout this paper we
pay much attention to nonomniscient BPCIs (abbreviated NBPCI). That is,
we will evaluate lab-wise coverage when the investigator's prior
distribution $\Pi$ differs from the PGD $P$.

It is worth noting that BPCIs have desirable properties from a
decision-theoretic point of view. The~situation is complicated in that
both coverage and length must be reflected in the loss function. Hence,
this function must be bivariate, or be a univariate combination of
coverage and length terms (which would necessitate some weighting of the
two). Robert (\citeyear{Ro1994}) gives some general discussion of this point. Despite
this complication, there are still results which link, and come close to
equating, BPCIs and admissible interval estimators (see, for instance,
Meeden and Vardeman, \citeyear{MeVa1985}). Thus, the common argument for Bayesian point
estimators having desirable frequentist properties does extend, albeit
with complications, to the case of interval estimators.

Additionally, there are large-sample results saying that in ``regular''
modeling situations with large sample sizes and priors with unrestricted
support, BPCIs will have frequentist coverage that converges to nominal
coverage, at every possible set of parameter values. These results are
based on obtaining a likelihood that dominates the prior given enough
data; as such, they are not very useful for our purposes, because later
we turn to problems in which no such domination occurs. We will however
find use for a variant of this result in which information is
accumulated over a sequence of studies. First, however, we illustrate
the operating characteristics of some interval estimators in a simple
but relevant situation.

\subsection{Example: Mixture of Near-Null and Important~Effects}

Say that $\theta$ represents the strength of a putative exposure-disease
relationship (which may indeed be one of a sequence of such
exposure-disease combinations to be investigated). For instance,
$\theta$ might be a risk difference or a log odds-ratio relating binary
exposure and disease variables. Suppose that $D$ is a univariate
sufficient statistic such that $D|\theta \sim N(\theta,\sigma^{2})$
where $\sigma^{2}$ is known. Then $( D \pm q_{\alpha /2}\sigma )$ can be
reported as a $100 \times( 1 - \alpha)\%$ frequentist confidence interval
(FCI) for $\theta$, where $q_{\alpha /2}$ is the $1 -\alpha /2$
standard normal quantile.

In the context of observational epidemiology, null or minimal effects
are common, and large effects are rare. Thus, the PGD giving rise to a
sequence of studies might have most of its mass at or near zero. For
instance, say the PGD is a mixture of two normal distributions: $N(
0,\varepsilon^{2} )$ with weight $p$ and $N( 0,k^{2}\varepsilon^{2} )$
with weight $1-p$, for a ``small'' $\varepsilon$ and $k > 1$. This
is interpreted as the first component generating minimal or near-null
associations, while the second gives rise to important as well as
near-null associations, for example, $| \theta_{i} | < 2\varepsilon$ and
$| \theta_{i} | > k\varepsilon$ might reasonably be described as
near-null and important respectively.

We simulate 500,000 parameter-data ensembles with $\varepsilon = 0.05$,
$p=0.85$, $k = 8$, and $\sigma^{2}=0.025$. If $\theta$ is a log odds-ratio,
then these values have $\exp(\theta_{i})$ within $(0.91,1.1)$ as near-null,
and $\exp(\theta_{i})$ outside $(0.67,1.5)$ as important. The~choice of
$\sigma^{2} = 2/((500)(0.2)(0.8))$ approximates the amount of
information for the log odds ratio when comparing two independent groups
(as in an unmatched case-control study) with 500 subjects per group and
exposure prevalences around 20\%.

The~first two rows of Table \ref{tab1} give operating characteristics of the FCI
and the (equal-tailed) OBPCI as interval estimators for $\theta$ (both
at the nominal 95\% level). Note that when used as the prior
distribution for $\theta$, the mixture distribution is conjugate, so
that computation of the OBPCI is straightforward. As is consistent with
theory, the labwise coverages of both procedures are within simulation
error of the nominal 95\%. On average, though, the OBPCI is considerably
shorter than the FCI, by almost a factor of two. This results from the
infusion of prior information.

\begin{table*}
\caption{Frequency properties of interval estimators based on 500,000
simulated parameter-data pairs from PGD with $\varepsilon = 0.05$, $p =
0.85$, $k = 8$. The~``omniscient'' posterior (OBPI)
uses these parameter values; the ``nonomniscient'' posterior
(NBPI) uses the values of $p$ and $k$ shown\protect\tabnoteref{t1}}
\label{tab1}
\begin{tabular*}{\textwidth}{@{\extracolsep{\fill}}lccccc@{}}
\hline
 & \textbf{Coverage \%} & \textbf{Avg. length} & \textbf{TDR \%} & \textbf{FDR \%} & \textbf{FNR \%}\\
 \hline
{FCI} & 95.0 & 0.62 & 6.3 & 17.0 & 11.5\\
{OBPI} & 95.0 & 0.33 & 2.2 & \phantom{0}0.6 & 14.1\\
{NBPI:} &  &  &  &  & \\
{$p=0.50$, $k=4$ } & 95.8 & 0.41 & 2.1 & \phantom{0}0.5 & 14.2\\
$p=0.50$, $k=12$ & 97.3 & 0.46 & 3.2 & \phantom{0}2.5 & 13.3\\
{$p=0.95$, $k=4$ } & 89.8 & 0.24 & 0.9 & \phantom{0}0.0 & 15.2\\
$p=0.95$, $k=12$ & 91.7 & 0.26 & 1.8 & \phantom{0}0.2 & 14.4\\
${N}(0,\nu^{2})$ & 94.8 & 0.44 & 2.1 & \phantom{0}0.6 & 14.1\\
${N}(0,0.5\nu^{2})$ & 92.1 & 0.36 & 0.8 & \phantom{0}0.0 & 15.2\\
${N}(0,2\nu^{2})$ & 96.5 & 0.51 & 3.5 & \phantom{0}3.4 & 13.0\\
\hline
\end{tabular*}
\tabnotetext[*]{t1}{Simulation standard errors for coverage, TDR $\approx 0.04\%$. The~simulation standard errors for FDR are considerably larger and variable,
since only a small portion (the TDR) of the simulated pairs contribute
to the estimated proportion.}
\end{table*}

Motivated by taking $| \theta| < 2\varepsilon$ as a minimal effect, we
also define the \textit{total discovery rate} (TDR), \textit{false
discovery rate} (FDR) and \textit{false nondiscovery rate} (FNR) for
interval estimation as follows: The~TDR is simply the proportion of
reported intervals that exclude the minimal range, that is, give
confidence that the effect is not minimal. The~FDR is then the
proportion of these ``discoveries'' that are false, that is, in which
the parameter actually does lie in the minimal range. Similarly, amongs
intervals intersecting the minimal range, the FNR is the proportion for
which the target is actually outside this range. We can then describe
how the OBPCI is more conservative than the FCI: The~OBPCI attains a
lower FDR at the cost of a higher FNR, as evidenced in the first two
rows of Table \ref{tab1}.

Investigators are not omniscient. To illustrate consequences of
defective prior information, we examine results in which the prior
distribution deviates from the PGD. Our example is far from a
comprehensive study of prior misspecification, and we doubt that such a
study could be done given all the contextual elements involved. Rather,
we wish to illustrate some qualitative points that will be relevant
later, regarding potential consequences of such misspecification.

Two sets of NBPCI results are given in Table \ref{tab1}. The~first set
corresponds to an investigator using the same form of a mixture-normal
prior with the correct value of $\varepsilon$ (which defines the notion
of a minimal effect and so is contextually established), but with
misspecified values of $p$ and $k$ (choosing $p=0.50$ or $p=0.95$, and
$k=4$ or $k=12$). The~second set corresponds to an investigator who
does not elucidate a mixture structure for the prior, but rather simply
applies a mean-zero normal prior. The~case where the prior variance
$\tau^{2}$ equals the PGD variance of $\nu^{2} = p\varepsilon^{2} +
(1-p)k^{2}\varepsilon^{2}$ is considered, as are the cases where the
prior variance is half/double the PGD variance. The~results in Table \ref{tab1}
underscore the disadvantage of the NBPCI relative to the FCI and the
unattainable OBPCI: The~labwise coverage now deviates from nominal.
Arguably, however, these deviations are modest. Moreover, the NBPCI tend
to maintain the other attractive features seen with the OBPCI, namely,
the much shorter average length and lower FDR compared to the FCI. Note
that the deviations from nominal coverage are less pronounced and tend
toward conservatism when the prior is more spread out than the PGD
($p=0.50$ in the first set of results, $\tau^{2}=2\nu^{2}$ in the
second). This is not surprising, since NBPCIs will resemble FCIs more
and more in the ``flat'' prior limit. In contrast, the deviations can be
markedly anticonservative when the prior is more concentrated ($p=0.95$
in the first set, $\tau^{2}=0.5\nu^{2}$ in the second). Thus, by using
very dispersed priors, we can improve the precision and reduce the FDR
of our intervals without incurring objectionable deviations from nominal
coverage. If, however, we ``get greedy'' and attempt to improve
performance by using overconfident priors, we risk unacceptable
deterioration of coverage.

In this and subsequent examples, we have used equal-tailed BPCIs because
these are intuitive and commonly reported. It is well known, however,
that for a given data set the HPD interval (or possibly region) is the
shortest interval with the specified Bayesian probability content. In
fact, Uno, Tian and Wei (\citeyear{UnTiWe2005}) prove an interesting result about labwise
coverage of HPD intervals, in the case that the prior and PGD coincide.
They show that the HPD interval with coverage $1-\alpha$ does not always
minimize \textit{average} width subject to obtaining labwise coverage
$1-\alpha$. Rather, the minimizing procedure in general involves the HPD
interval with coverage $1-\alpha(D)$, such that
$E\{\alpha(D)\}=\alpha$, where the data-dependent coverage level
$\alpha(D)$ arises by thresholding the posterior densities for all
studies at the same cutoff value. Thus, in cases where the width of the
posterior density varies across studies, HPD intervals with higher
(lower) coverage levels will be reported for studies with narrower
(wider) posterior densities. While we do not pursue this further here,
it is worth emphasizing that the simple and intuitive interpretations
associated with using a BPCI of fixed coverage level can be sacrificed
in order to obtain intervals which are narrower on average.

Given the performance issues illustrated in Table \ref{tab1}, we find it overly
simplistic to argue against the use of Bayesian interval estimation
simply because prior specification is required, and because it is
impossible to get this specification exactly right in terms of matching
the PGD. Furthermore, as we discuss in the next section, by using a
hierarchically structured prior, one can effectively move the prior
closer to the PGD as a sequence of studies unfolds.

\section{Hierarchical Prior Distributions}\label{sec3}

As we have emphasized, consideration of a sequence of studies is a
conceptual device capturing the reality facing most investigators.
Studies in\break medicine and public health are nowhere near identical in
design, conduct and population studied and, hence, there is no basis for
asserting parameter equality across these studies. This fact is the
rationale for random-effects models in meta-analysis, which typically
employ very simple models for the PGD. When a new study is performed,
however, data from $m$ previous studies can be used to improve the prior
distribution for $\theta$ by using a hierarchically structured prior
distribution, which in turn will make the labwise coverage closer to
nominal (as $m$ increases, particularly). This further strengthens the
frequentist appeal of Bayesian intervals.

Say that the study to be carried out has parameter-data ensemble
$(D,\theta)$ and is preceded by $m$ earlier studies with ensembles $(
D_{1}^{*},\theta_{1}^{*} ), \ldots,( D_{m}^{*},\theta_{m}^{*} )$. If the
$m+1$ ensembles are independent and identically distributed (i.e., each
according to the PGD and the data model), it makes sense to allow the
interval estimator for $\theta$ to depend on the earlier data as well as
the current data. The~labwise coverage (\ref{for1}) then generalizes to
\begin{equation}\label{for2}
C(I,P) = \operatorname{Pr}\{\phi \in I(D;D^{*})\},
\end{equation}
where $D^{*} = (D_{1}^{*},\ldots,D_{m}^{*})$ and the probability is taken
with respect to the joint distribution of the $m+1$ parameter-data
ensembles.

The~standard Bayesian approach to borrowing\break  strength across studies
involves a hierarchical prior. That is, the prior $\Pi$ asserts that the
$m +1$ components of $(\theta^{*},\theta)$ are independent and
identically distributed given a further parameter vector $\lambda$. Then
$\lambda$ itself is assigned a prior distribution. Application of Bayes
theorem to form the posterior distribution on $\theta$ involves the
likelihood contribution of $D|\theta$, with $\Pi(\theta\vert D^{*}) =
\int\Pi(\theta\vert\lambda)\Pi(\lambda\vert D^{*})\,d\lambda$ playing the
role of the prior. That is, the earlier studies inform the value of
$\lambda$, which in turn informs $\theta$, in advance of observing $D$.
If the PGD is well approximated by the posited $(\theta\vert\lambda)$
prior for some value of $\lambda$ (say, $\lambda_{0})$, and if the
number of previous studies $m$ is large, then $\Pi(\lambda\vert D^{*})$
should be concentrated near $\lambda_{0}$. Thus, the ``effective
prior''
being applied to $\theta$ will be close to the PGD, which should result
in labwise coverage for BPCIs that is close to nominal.

The~use of hierarchical priors to ``borrow strength'' across studies and
the evaluation of coverage along the lines of (\ref{for2}) originates under the
rubric of ``empirical Bayes'' procedures (see, for instance, Morris, \citeyear{Mo1983}),
which typically involve a non-Bayesian approximation to $\Pi(\lambda\vert
D^{*})$. With the advent of better algorithms and machines for Bayesian
computation, however, fully Bayesian ``hierarchical modeling'' is now
commonplace. It should also be noted that treating the parameter values
for the $m+1$ studies as exchangeable as described is a modeling
assumption that will sometimes be inappropriate. Notably, in a situation
where all studies focus on the same relationship at different calendar
times, the assumption may be dubious but may be weakened to allow for
trends. For instance, the ``Ty Cobb'' example in Morris (\citeyear{Mo1983}) involves
explicit modeling of a time trend for parameter values corresponding to
consecutive calendar years. Analogously, the assumption may be weakened
to allow for group effects; for
example, the occupational-cancer example in Greenland (\citeyear{Gr1997}) explicitly
models changes in association over cancer type and exposure type. In
such examples it is a residual component of the study parameters (after
time or group effects are regressed out) that is assumed exchangeable.
The~exchangeability inherent in assuming the original study parameters
are conditionally i.i.d. would be appropriate only if no important
information is conveyed by the time order or other known and varying
characteristic of the parameters being modeled as an ensemble.

To give a simple illustration, suppose again that $D|\theta \sim
N(\theta,\sigma^{2})$ with $\sigma^{2}$ known. For computational ease
we first consider a simpler PGD than previously, namely, a
$N(\lambda_{0},\tau^{2})$ distribution. Consider a partially omniscient
investigator who knows the variance of the PGD, but not the mean, and in
hierarchical fashion assigns the prior $\theta\vert\lambda \sim
N(\lambda,\tau^{2}), \lambda \sim\break  N( \delta,\omega^{2} )$. The~marginal
prior distribution is then $\theta \sim
N(\delta,\tau^{2}+\omega^{2})$. In the absence of previous studies, or
with an i.i.d. prior assuming independence of $\theta$ and~$\theta^{*}$, the posterior on $\theta$ would arise from combining this
prior with $D$, and the discrepancy between this prior and the PGD would
induce some degree of non-nominal labwise coverage. With a correct
hierarchical prior, however, the previous data will pull
$\Pi(\lambda\vert D^{*})$ toward $\lambda_{0}$, and hence pull
$\Pi(\theta\vert D^{*})$ toward the PGD.

\begin{figure}[b]

\includegraphics{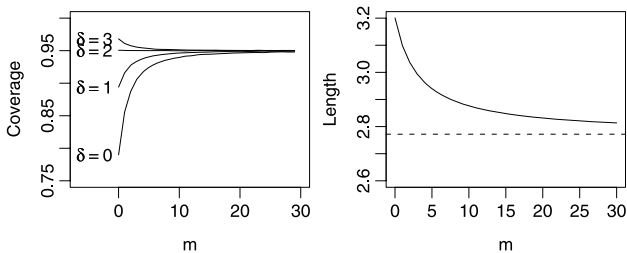}

\caption{Labwise coverage and interval length for the nominal 95\% BPCI, as a function of the number
of previous studies $m$.  Coverage is given for four choices of hyperparameter $\delta$, whereas the length does not depend on $\delta$.
The~dashed horizontal line in the second panel corresponds to the length of the OBCPI.}\label{fig1}
\end{figure}

The~present setting is sufficiently simple that the LWC given in (\ref{for2}) can
be computed directly (one-dimensional numerical integration is required,
but repeated simulation of data is not). As an example, suppose $\sigma
=1$, and the PGD has $\lambda_{0} = 3$, $\tau = 1$. Consider four prior
distributions for $\lambda$, with $\omega =1$, and $\delta = 0, 1, 2,
3$. Note that these priors range from very bad (mean of the PGD lies
three prior standard deviations away from the prior mean) to
unrealistically good (mean of the PGD coincides with the prior mean).
Figure \ref{fig1} illustrates the coverages (\ref{for2}) for the resulting 95\% BPCIs, as
the number of previous studies $m$ increases. When $m=0$ (thought of as
either no previous studies, or as an i.i.d. prior across studies),
the coverage ranges from less than 80\% for the ``worst'' prior to
somewhat above 95\% for the ``best'' prior. As Figure~\ref{fig1} illustrates,
however, for all the priors the coverage converges quite quickly to the
nominal 95\% as $m$ increases, to match the OBPCI coverage. The~figure
also displays the interval width as a function of $m$. In this simple
setting the width is governed by
\begin{eqnarray}\label{for3}
&&\operatorname{Var}(\theta\vert D,D^{*})\nonumber\\
 &&\quad =
\sigma^{2}\tau^{2}(\sigma^{2}+\tau^{2})^{-1}\nonumber\\[-8pt]\\[-8pt]
&&\qquad {}\times [1 +
\sigma^{2}\omega^{2}\tau ^{-2}\{(m+1)\omega^{2} \nonumber\\
&&\hspace*{78pt}\qquad {}+ \sigma^{2} +
\tau^{2}\}^{-1}],\nonumber
\end{eqnarray}
which depends on neither the observed data nor the hyperparameter
$\delta$. Note that by $m=30$ most of the potential reduction in width
has been realized, that is, the width is close to the OBPCI width which
corresponds to the $m \rightarrow \infty$ limit of (\ref{for3}). While the
convergence of coverage and length to match the OBPCI represents a
well-known calibration feature of Bayes and empirical-Bayes estimation,
it is interesting to see how rapidly it can proceed in simple settings.

Of course, the above illustration is very simplistic, particularly as
the variance components involved in the prior are taken to be known, and
only the mean of the PGD must be learned via previous data. At the other
end of the spectrum, one anticipates that complex PGDs, such as those
involving a mixture of near-null and important effects, will require a
larger number of studies before they are estimated well. To investigate
this, we reconsider the example of the previous section, involving a
mixture of near-null and important effects. Now, however, we treat
$(p,k)$ as unknown parameters with prior distributions $p \sim
\operatorname{Uniform}(0,1)$ and $k \sim \operatorname{Uniform}(4,20)$. As before, the PGD is based on
$p=0.85$ and $k=8$. Empirical results on coverage and average length
appear in Table~\ref{tab2}. These are based on only 1000 simulated
parameter-data meta-ensembles (with each meta-ensemble encompassing
$m+1$
parameter-data ensembles), since the posterior computation is
burdensome. In particular, a~simple Markov chain Monte Carlo algorithm
(with random walk proposals) is applied to sample from $(p,k|D,D^{*})$,
while $(\theta\vert p,k,D,\break D^{*})=(\theta\vert p,k,D)$ can be sampled
from directly. Results for coverage are quite appealing, in that the
labwise coverage (\ref{for2}) modestly exceeds nominal when the number of
previous studies $m$ is small, presumably because
$\Pi(\theta\vert D^{*})$ is very flat, and also modestly exceeds
nominal when $m$ is large, presumably because $\Pi(\theta\vert D^{*})$
is close to the PGD. However, the very slow convergence of
$\Pi(\theta\vert D^{*})$ to the PGD is manifested by the average
interval width. Even with $m=100$ previous studies, the average width is
still 44\% larger than that of the OBCI. Nonetheless, it is 23\%
narrower than the FCI, a~worthwhile gain paid for by a minor
conservatism.

\begin{table}[b]
\caption{Empirical properties of interval estimators with a prior
distribution on $(p,k)$ and $m$ previous studies. Results based
on~1000 simulated parameter-data meta-ensembles}
\label{tab2}
\begin{tabular*}{\columnwidth}{@{\extracolsep{\fill}}lcc@{}}
\hline
$\bolds{m}$  & \textbf{Coverage \%} & \textbf{Avg. length}\\
\hline
\phantom{00}0 & 96.4 & 0.515\\
\phantom{0}10 & 95.9 & 0.491\\
\phantom{0}20 & 95.8 & 0.487\\
100 & 95.4 & 0.476\\
\hline
\end{tabular*}
\end{table}

There are many examples of hierarchical modeling in the literature,
where unknown means and variances are themselves modeled via prior
distributions or estimated via marginal likelihood. These methods have
performed quite well in large-scale simulations and in applications that
provide subsequent validation (Brown, \citeyear{Br2008}). Special methodology for
inference about the distribution of a sequence of effects has expanded
apace, driven by work on multiple comparisons (and particularly false
discovery rates) in genome studies (see, for instance, Efron et.
al., \citeyear{EfTiStTu2001} and Newton and Kendziorski, \citeyear{NeKe2003}).

We have emphasized that \textit{formal} inclusion of previous studies on
various phenomena within a research team's domain of study can have
positive benefits for subsequent studies within this domain, in terms of
both labwise coverage and average width. Consequently, a~formal scheme
to obtain a prior which is close to the PGD for a given domain seems
desirable when practical. In other circumstances, however, it should be
\mbox{possible} to \textit{informally} use previous studies in constructing a
reasonable prior distribution. As alluded to earlier, for instance, in
many sub-fields of epidemiology investigators do have well-grounded
notions concerning the across-study prevalence of near-null effects and
magnitude of important effects. One anticipates that a prior formed from
direct elicitation of the investigators' views\break should not deviate
greatly from a prior formed from formally updating a ``flat'' prior based
on previous studies. Regardless of which route is taken, construction of
a prior which is reasonably close to the PGD for future studies in the
domain seems to be a realistic and worthwhile goal. With this
encouraging message in hand, we now turn to examining the use of
Bayesian interval estimators in nonidentified model settings.

\section{Interval Estimation in Nonidentified~Models}\label{sec4}

The~case for BPCIs versus FCIs seems mixed thus far, particularly as
FCIs are guaranteed to have correct labwise coverage, without requiring
any knowledge of the PGD. But in a large class of statistical problems,
construction of valid FCIs is not possible. Recall that in general a
model is nonidentified if there are multiple sets of parameter values
giving rise to the same distribution of observables. We have argued that
this class of models is the only realistic choice in most observational
studies of human health and society (Greenland, \citeyear{Gr2005}; Gustafson, \citeyear{Gu2006}).
This is particularly true in disciplines such as epidemiology where
honest appraisal of what modeling assumptions are justified, and what
limitations are inherent in the available data, \textit{ought} to lead
investigators to nonidentified models routinely.

Identifiable models are desirable when they can supply root-$n$ consistent
estimators of target parameters, as in classic industrial and laboratory
experiments. With study problems such as measurement error, missing
data, selection bias and unmeasured confounders, however, extremely
strong assumptions may be required to attain an identified model. Most
statistical methods assume absence of such problems, and the remainder
assume that the form of the problems is known up to a few identifiable
parameters. Either way, there is a strong possibility that the resulting
model is grossly misspecified, with the resulting FCIs exhibiting
excessive precision and severe undercoverage for the inferential target.

Put another way, using an overly simplified model for the sake of
identifiability results in root-$n$ consistent inference \textit{for the
wrong parameter} (e.g., an unconditional association, when the desired
inferential target is an association conditional on an unmeasured
covariate) (Greenland, \citeyear{Gr2003}, \citeyear{Gr2005}; Gustafson, \citeyear{Gu2006}). If, as usual, the
parameter being estimated does not equal the target parameter, the
interval coverage for the latter will tend to zero as the sample size
increases.

Backing away from untenable assumptions may result in a model that is
better specified (closer to reality, or at least better representing the
true inferential target), but which lacks identifiability. There is
extreme hesitance among statisticians regarding the use of nonidentified
models, because they do not give rise to estimators with familiar
statistical properties, such as \mbox{root-$n$} shrinkage of interval estimators
\textit{to some value}. But for Bayesian analysis there is no conceptual
or computational difference in how inferences are obtained from a
nonidentified model compared to an identified model. In fact, from a
radical subjective Bayesian perspective, identification is a matter of a
degree and always a function of the full prior (including the prior for
the data given the parameters).

In summary, in nonidentified problems there is no route to FCIs
achieving exactly nominal coverage for any set of underlying parameter
values. If in these settings we simplify the model to the point of
identifiability, then FCIs are readily obtained via standard methods,
but are likely to have grossly incorrect coverage probabilities due to
misspecification. Without simplification, models are nonidentified,
which precludes construction of FCIs having the nominal coverage
probability at every point in the parameter space.

Some frequentist approaches to problems of this sort involve (i)
specifying bounds (rather than prior distributions) on key parameters,
and (ii) constructing interval estimators having \textit{at least}
nominal coverage at every point in the parameter space, with the
consequence that the coverage will be higher than nominal at most
parameter values. Some recent suggestions along these lines include
Imbens and Manski (\citeyear{ImMa2004}), Vansteelandt et al. (\citeyear{VaGoKeMo2006}) and Zhang (\citeyear{Zh2009});
we illustrate such an approach in the first of the two examples below.
Conversely, the use of Bayesian or approximately Bayesian inferences
from nonidentified models was suggested at least as far back as Leamer
(\citeyear{Le1974}), and has long been discussed under special topics such as
nonignorable missingness (Little and Rubin, \citeyear{LiRu2002}). It has also attracted
considerable attention in recent literature; see, for instance,
Dendukuri and Joseph (\citeyear{DeJo2001}); Greenland (\citeyear{Gr2003}, \citeyear{Gr2005}); Gustafson (\citeyear{Gu2005b});
Gustafson and\break Greenland (\citeyear{GuGr2006a}, \citeyear{GuGr2006b}); Hanson, Johnson
and\break Gardner (\citeyear{HaJoGa2003});
Joseph, Gyorkos and Coupal (\citeyear{JoGyCo1995}); McCandless, Gustafson and Levy (\citeyear{McGuLe2007},
\citeyear{McGuLe2008}); Scharfstein, Daniels and Robins (\citeyear{ScDaRo2003}).

For Bayesian procedures, the exact attainment of nominal labwise
coverage by an OBPCI still holds under nonidentified models. The~result
in general (for any kind of model) is known, but surprisingly
unemphasized in the literature (see Rubin, \citeyear{Ru1984}, and Rubin and Schenker,
\citeyear{RuSc1986}, for exceptions). Yet it seems to be a useful reference point, as
it provides a clear calibration, or ``anchor,'' for an interval
estimation procedure in a nonidentified model. On the other hand, we
generally expect the choice of prior to be far more influential on the
posterior distribution when the model is nonidentified, so that lab-wise
coverage may deviate rapidly from nominal as the prior distribution
deviates from the PGD. We investigate this phenomenon in the two
examples below.

\subsection{Example: Prevalence Survey with~Nonresponse}

Vansteelandt et al. (\citeyear{VaGoKeMo2006}) illustrate some frequentist
techniques for sensitivity analysis in nonidentified models in the
following setting. A binary outcome $Y$ may be observed ($R=1$) or
missing ($R=0$, nonresponse) for each study unit, so that the available
data consist of $n$ i.i.d. realizations of $(RY, R)$. The~inferential target is the outcome prevalence, $\pi = \operatorname{Pr}(Y=1)$, while
the missingness may be informative, that is, $Y$ and $R$ may be
associated. One parameterization for this situation is $p = \operatorname{Pr}(R=1)$, $s
= \operatorname{Pr}(Y=1|R=1)$, and $\gamma = \operatorname{logit}\{\operatorname{Pr}(Y=1|R=0)\} - \theta$ where
$\theta = \operatorname{logit}(s)$. Then the inferential target is $\pi =
(1-p)\operatorname{expit}(\theta +\gamma) + ps$. This is a nonidentified inference
problem because the likelihood for the observed data depends only on $p$
and $s$, while the inferential target also depends on~$\gamma$.

We consider the coverage and average length of three interval estimators
for $\pi$. The~first is the na\"{\i}ve interval estimator obtained by
assuming $\gamma = 0$, that is, assuming the missingness is completely
at random, and estimating $\pi$ as the sample proportion of the observed
outcomes. The~second is an interval estimator suggested by Vansteelandt
et al. (\citeyear{VaGoKeMo2006}), designed to have at least nominal frequentist
coverage (approximately) under every fixed value of $\gamma$ in a
specified interval $I$; we take $I = (-2,2)$ in the present example. Let
$\hat{\pi} _{l}$ and~$\hat{\pi} _{u}$ be the estimates of $\pi$ when
fixing the value of $\gamma$ at the lower and upper endpoints of $I$
respectively. Then the interval estimator with target level $1 - \alpha$
is of the form $( \hat{\pi} _{l} - q_{\alpha ^{*}/2}se( \hat{\pi} _{l}
),\hat{\pi} _{u} + q_{\alpha ^{*}/2}se( \hat{\pi} _{u} ) )$, where
$\alpha^{*}$ is chosen to make the minimum coverage as $\gamma$ varies
in $I$ equal to $1 - \alpha$ (with the minimum attained at one of the
endpoints). We refer to this interval as a conservative frequentist
confidence interval (CFCI). The~relationship between $\alpha^{*}$ and
$\alpha$ depends on the unknown parameters, hence, estimates are plugged
in and the coverage properties become approximate rather than exact.
Vansteelandt et al. (\citeyear{VaGoKeMo2006}) call interval estimators of this
form ``pointwise estimated uncertainty regions,'' since the coverage
claim applies to the true value of the target parameter. These authors
also propose ``weak'' and ``strong'' estimators with coverage claims
pertaining to the set of all target parameter values consistent with the
observed data law (i.e., interval estimation of an interval). For
more details see Vansteelandt et al. (\citeyear{VaGoKeMo2006}).

The~third interval estimator is the equal-tailed Bayes\-ian credible
interval arising from a uniform prior distribution for $\gamma$ on the
same interval $I$, along with $\operatorname{uniform}(0,1)$ priors for both $p$ and
$s$. Under this specification the parameters $p$, $s$ and $\gamma$
remain independent of one another a posteriori, with beta
distributions for $p$ and $s$ arising from binomial updating, and a
uniform posterior distribution on $I$ for $\gamma $; that is, no
updating of $\gamma$ occurs.

Empirical labwise coverage and average length for nominal 95\% intervals
are given in Table \ref{tab3}. The~PGDs used have normal distributions for $\beta
= \operatorname{logit}(p)$ and $\theta = \operatorname{logit}(s)$ with $\mu_{\beta} = \operatorname{logit} 0.67$,
$\sigma_{\beta} = (\operatorname{logit} 0.89 - \operatorname{logit} 0.67)/2$, $\mu_{\theta} = \operatorname{logit}
0.5$ and $\sigma_{\theta} = (\operatorname{logit} 0.8 -\break \operatorname{logit} 0.5)/2$. Thus, the PGD
for $(p,s)$ concentrates around more typical-use scenarios than
does the prior for these parameters. The~PGD is completed by $\gamma
\sim \operatorname{uniform}(J)$ for various specifications of interval \textit{J.}
Note that one specification is the single-point interval $J=[2,2]$, which
corresponds to fixing $\gamma$ at the endpoint of $I$, and hence
corresponds to a partially frequentist evaluation of coverage. Note also
that the average interval lengths do not depend on the specification of
$J$ for this problem, since the distribution of the observed data (under
the joint distribution of parameters and data) does not depend on $J$.
Thus, the average lengths of 0.11 for the na\"{\i}ve interval, 0.33 for
the CFCI, and 0.28 for the BPCI apply for any $J$.

\begin{table}
\caption{Empirical coverage probabilities and average lengths for
nominal 95\% interval estimators of a prevalence $\pi$. Results are
given for na\"{\i}ve estimator, the CFCI and the BPI. For each choice of
PGD (i.e., choice of interval $J)$, results are based on 10,000
simulated parameter-data ensembles with a sample size of $n=500$.
Simulation standard errors for coverages are 0.5\% or less. Both the
CFCI and the BPI assume an interval range $I=(-2,2)$ for $\gamma$}
\label{tab3}
\begin{tabular*}{\columnwidth}{@{\extracolsep{\fill}}lccc@{}}
\hline
$\bolds{J}$ \textbf{in PGD:} & \textbf{Na\"{\i}ve} & \textbf{CFCI} & \textbf{Bayes}\\
\hline
$J=(-2,2)$ & 42\% & \phantom{0}99\% & \phantom{0}95\%\\
$J=(-3,3)$ & 30\% & \phantom{0}93\% & \phantom{0}80\%\\
$J=(-1,1)$ & 67\% & 100\% & 100\%\\
$J=(-1,3)$ & 40\% & \phantom{0}95\% & \phantom{0}84\%\\
$J=(2,2)$ & \phantom{0}9\% & \phantom{0}95\% & \phantom{0}71\%\\
Average length & 0.11 & 0.33 & 0.28\\
\hline
\end{tabular*}
\end{table}

Table \ref{tab3} verifies that when $J$ in the PGD and $I$ in the prior coincide,
the Bayesian intervals have LWC within simulation error of nominal,
despite the discrepancy between the uniform priors for $(p,s)$
and the logit-normal PGDs for $(p,s)$. In contrast, the CFCI
approach is indeed quite conservative when $I$ and $J$ coincide, with
labwise coverage of 99\% and average length 17\% greater than the BPCI.
As expected, the labwise coverage of both the CFCI and the BPCI is
highly affected by any discrepancy between $I$ and $J$. As advertised,
the CFCI achieves conservative coverage in all cases, except for a
slight dip below nominal in the case that $J$ is wider than (and
contains) $I$. Note, in particular, that the CFCI achieves nominal
coverage when $\gamma$ is fixed at an endpoint of \textit{I,} whereas
the BPCI coverage drops to 71\% in this setting.

The~differences between labwise coverage of BPCIs and CFCIs are somewhat
hidden in Table \ref{tab3}, since nominal 95\% intervals do not have much
``room'' to obtain higher than nominal coverage. Thus, we also report
results for nominal 80\% intervals (Table \ref{tab4}). Admittedly, such intervals
are seldom reported in practice (though see Greenland et al.,
\citeyear{GrShKaPoKe2000}, for an exception), but they are useful for gauging the extent to
which a given interval estimator is conservative. The~average lengths of
these intervals are 0.069 (na\"{\i}ve), 0.29 (CFCI) and 0.22 (BPCI).
When $I$ and $J$ match, we now see very substantial over-coverage (96\%)
for the CFCI, with an average width 30\% greater than for the OBPCI. We
also see more clearly the over-coverage that results for both CFCI and
BPCI when $J$ is narrower than $I$.

\begin{table}
\caption{Empirical coverage probabilities and average lengths for
nominal 80\% interval estimators of a prevalence $\pi$. Both the CFCI
and the BPI assume an interval range $I=(-2,2)$ for $\gamma$. The~table
entries are as per Table \protect\ref{tab2}}
\label{tab4}
\begin{tabular*}{\columnwidth}{@{\extracolsep{\fill}}lccc@{}}
\hline
$\bolds{J}$ \textbf{in PGD:} & \textbf{Na\"{\i}ve} & \textbf{CFCI} & \textbf{Bayes}\\
\hline
$(-2,2)$ & 27\% & \phantom{0}96\% & 80\%\\
$(-3,3)$ & 19\% & \phantom{0}83\% & 59\%\\
$(-1,1)$ & 47\% & 100\% & 98\%\\
$(-1,3)$ & 26\% & \phantom{0}87\% & 69\%\\
$(2,2)$& \phantom{0}4\% & \phantom{0}80\% & 31\%\\
Average length & 0.069 & 0.29 & 0.22\\
\hline
\end{tabular*}
\end{table}

The~BPCI and the CFCI are constructed to satisfy different criteria, and
we are not attempting to argue than one is better than the other. In
particular, note the tradeoff exhibited in Tables \ref{tab3} and \ref{tab4}. If the
investigator has an interval of values $I$ in mind for $\gamma$, then
the CFCI has a conservatism which may be appealing: at least nominal
coverage can be obtained with respect to any averaging across values in
$I$, including the selection of single points. On the other-hand,
if labwise coverage with respect to the $\operatorname{Uniform}(I)$ distribution is at
issue, then the BPCI will be shorter on average, and have correct
coverage. We do emphasize that this correct coverage constitutes a
calibration property of the BPCI which the CFCI does not possess. That
is, without doing simulation, we do not know to what extent the CFCI
based on interval $I$ will exhibit higher than nominal labwise coverage
when the PGD is based on $I$. But we do know automatically that the BPCI
using $I$ in the prior will exhibit correct labwise coverage when the
PGD is based on $I$. Thus, the BPCI is anchored via the investigator's
knowledge that \textit{exactly} nominal coverage \textit{would} be
obtained in a sequence of studies with PGD equal to the prior, and
presumably \textit{at least} nominal coverage \textit{would} eventually
be attained in a sequence of studies in which the support of the prior
contains the PGD. In this sense, posterior coverage is conservative
precisely when the prior is conservative relative to the PGD. The~CFCI
labwise coverage has a more murky connection to the PGD, which is the
price it pays for obtaining correct frequentist coverage at the
endpoints of the prior interval $I$.

\subsection{Example: Case-Control Study with~Misclassification}

Consider an unmatched case-control study of the association of a disease
indicator $Z$ and a binary exposure indicator $X$, with $X$ subject to
independent nondifferential misclassification. Let $r_{0} =\break
\operatorname{Pr}(X=1|Z=0)$ and $r_{1} = \operatorname{Pr}(X=1|Z=1)$ be the prevalences of actual
exposure among nondiseased and diseased source population members, and
let $\mathit{SN} = \operatorname{Pr}(X^{*}=1|X=1)$ and $\mathit{SP} = \operatorname{Pr}(X^{*}=0|X=0)$ be the
sensitivity and specificity of the exposure classification in the study.
The~numbers \textit{apparently} exposed among the $n_{0}$ nondiseased
controls and $n_{1}$ diseased cases in the study are modeled as
$Y_{i}\sim \operatorname{Bin}( n_{i},\theta_{i} )$ for $i=0$ and $i=1$ respectively,
with $\theta_{i} = r_{i}\mathit{SN} + ( 1 - r_{i} )( 1 - \mathit{SP} ) =
\operatorname{Pr}(X^{*}=1|Z=i)$. If all four parameters $( r_{0},r_{1},\mathit{SN},\mathit{SP} )$ are
unknown, then this model is not identified by the observed counts
($y_{1}, y_{0}, n_{1}-y_{1}, n_{0}-y_{0})$. Bayesian inference under
this model is considered by Gustafson, Le and Saskin (\citeyear{GuLeSa2001}), Gustafson
(\citeyear{Gu2003}), Greenland (\citeyear{Gr2005}), Chu et al. (\citeyear{ChWaCoGr2006}) and Gustafson and
Greenland (\citeyear{GuGr2006a}), among others.

We consider prior distributions and PGDs of the following form: A
bivariate normal distribution for the logit prevalences $(\operatorname{logit}r_{0},
\operatorname{logit}r_{1})$, with correlation $\rho$ and identical marginals (mean
$\mu$ and variance $\tau^{2})$. The~log-odds ratio, $\beta =
\operatorname{logit}(r_{1}) - \operatorname{logit}(r_{0})$, is then distributed as $N\{ 0,( 1 -
\rho)2\tau^{2} \}$. The~correlation is essential to reflect the fact
that information about the exposure prevalence in one group would alter
bets about the prevalence in the other group, due to prior information
about $\beta$ (Greenland, \citeyear{Gr2001}). $\mathit{SN}$ and $\mathit{SP}$ are here
taken as independent of the exposure prevalences and each other,
however, with $\mathit{SN}\sim \operatorname{Beta}( a_{N},b_{N} )$ and $\mathit{SP}\sim \operatorname{Beta}( a_{P},b_{P}
)$; more realistic priors might allow dependent $\mathit{SN}$ and
$\mathit{SP}$ (Chu et al., \citeyear{ChWaCoGr2006}; Greenland and Lash, \citeyear{GrLa2008}), or one could
instead reparameterize the problem to make prior independence reasonable
(Greenland, \citeyear{Gr2009}).

Bayesian computation is readily implemented via the efficient algorithm
of Gustafson, Le and Saskin (\citeyear{GuLeSa2001}). While this algorithm takes advantage
of structure imbued by assigning uniform priors on prevalences, we can
use importance sampling to adapt the algorithm output to the present
prior specification. As an example, $m = 10{,}000$ parameter-data
ensembles with $n_{1} = n_{2} = 500$ are drawn from the PGD based on $\mu
= - 2.3$, $\tau = 1.17$, $\rho = 0.76$, $a_{N} = a_{P} = 18$, $a_{N} = a_{P} =
4$. These choices produce a 95\% logit-symmetric interval for each
$r_{i}$ of $(0.01, 0.50)$ and a 95\% log-symmetric interval for
$e^{\beta}$ of $(0.2, 5.0)$. Also, the modes of the $\mathit{SN}$ and
$\mathit{SP}$ distributions are 0.85, with 95\% logit-symmetric intervals
of $(0.637, 0.946)$.

For each data set, seven interval estimates for $\beta$ are constructed:

\begin{longlist}[(iii)]
\item[(i)] the standard FCI assuming no misclassification;

\item[(ii)] an FCI derived by taking $\mathit{SN}=0.85$ and\break
\mbox{$\mathit{SP}=0.85$} as known values;

\item[(iii)] the omniscient BPCI arising when the prior distribution
coincides with the PGD;
\end{longlist}

Nonomniscient BPCIs with priors based on correct specification of $(
\mu,\tau,\rho)$ but:

\begin{longlist}[(vii)]
\item[(iv)] $a_{N} = a_{P} = 9.5$, $b_{N} = b_{P} = 2.5$ (keeping the
prior modes on $\mathit{SN}$ and $\mathit{SP}$ at 0.85 but making the
distribution more diffuse);

\item[(v)] $a_{N} = a_{P} = 26.5$, $b_{N} = b_{P} = 5.5$ (modes at 0.85 but
overly concentrated);

\item[(vi)] $a_{N} = a_{P} = 23.5$, $b_{N} = b_{P} = 8.5$ (still overly
concentrated and modes shifted down to 0.75);

\item[(vii)] $a_{N} = a_{P} = 28.5$, $b_{N} = b_{P} = 3.5$ (still overly
concentrated and modes shifted up to 0.95).
\end{longlist}
Empirical properties of the interval estimators (at the nominal 95\%
level) are described in Table \ref{tab5}.

\begin{table}[b]
\caption{Empirical labwise properties of nominal 95\% interval
estimators for a log odds ratio $\beta$ based on 10,000 simulated
parameter-data ensembles. The~simulation standard errors for coverage
are less than 0.5\%. Results for estimator (ii) are based only on the
81\% of~ensembles for which the method works}
\label{tab5}
\begin{tabular*}{\columnwidth}{@{\extracolsep{\fill}}lcc@{}}
\hline
 & \textbf{Coverage} & \textbf{Avg. length}\\
\hline
(i)-FCI & 44\% & 0.60\\
(ii)-FCI & 81\% & 2.20\\
(iii)-OBPI & 95\% & 2.02\\
(iv)-NBPI & 95\% & 2.12\\
(v)-NBPI & 94\% & 1.94\\
(vi)-NBPI & 95\% & 2.32\\
(vii)-NBPI & 87\% & 1.56\\
\hline
\end{tabular*}
\end{table}

\begin{table*}[b]
\caption{Near-frequentist coverage in the case-control study with
misclassification example}
\label{tab6}
\begin{tabular*}{\textwidth}{@{\extracolsep{\fill}}lccccc@{}}
\hline
 & $\bolds{\mathit{SP}^*=0.63}$ & $\bolds{\mathit{SP}^*=0.77}$ & $\bolds{\mathit{SP}^*=0.83}$ & $\bolds{\mathit{SP}^*=0.88}$ & $\bolds{\mathit{SP}^*=0.95}$\\
\hline
$\mathit{SN}^*=0.63$ & 90\% & 95\% & 97\% & 97\% & 98\%\\
$\mathit{SN}^*=0.77$ & 92\% & 98\% & 99\% & 99\% & 99\%\\
$\mathit{SN}^*=0.83$ & 93\% & 99\% & 99\% & 99\% & 99\%\\
$\mathit{SN}^*=0.88$ & 92\% & 99\% & 99\% & 99\% & 99\%\\
$\mathit{SN}^*=0.95$ & 93\% & 98\% & 99\% & 99\% & 98\%\\
\hline
\end{tabular*}
\legend{NOTE: Evaluation is for parameter values $\theta^{*}$ given by
$r^*=(0.10, 0.15)$ and the indicated values of
$(\mathit{SN}^*,\mathit{SP}^*)$, using $\alpha =0.01$ of the $m=100{,}000$ simulated
parameter-data ensembles in each instance. The~chosen values for
$(\mathit{SN}^*,\mathit{SP}^*)$ correspond to 2.5{th}, 25{th},
50{th}, 75{th} and 97.5{th} percentiles of the prior
distribution.}
\end{table*}

In the previous example, the joint posterior density was a product of
the marginal posterior density for the two parameters appearing in the
likelihood function and the marginal posterior density (equal to the
prior density) for the one parameter not in the likelihood. This
factorization simplified the mathematics of how the prior influences the
posterior distribution of the target parameter. In the present example,
however, the structure of the problem is more nuanced. As emphasized by
Gustafson, Le and Saskin (\citeyear{GuLeSa2001}), the support of the two parameters not
in the likelihood, $(\mathit{SN}, \mathit{SP})$, depends on the values of the two
parameters in the likelihood, $( \theta_{0},\theta_{1} )$, since by
construction $1-\mathit{SP}$ and $\mathit{SN}$ must straddle both
$\theta_{i}$ values. To some extent then, the posterior distribution of
$(\mathit{SN}, \mathit{SP})$ can depend on the data, even though these parameters do
not appear in the likelihood function. Gustafson (\citeyear{Gu2005a}) discusses such
\textit{indirect learning} about parameters in nonidentified models in
more general terms.

Given that the data can provide some information about $(\mathit{SN}, \mathit{SP})$,
one might anticipate that the NBPCI coverage would be less sensitive to
the choice of prior than in a situation without any indirect learning.
The~results in Table \ref{tab5} bear this out, with the coverage of nominal 95\%
NBPIs ranging from 87\% to 95\% across the priors considered. In accord
with theory, the OBPCI coverage is within simulation error of nominal,
which can be regarded as a check that our scheme for posterior
computation is working adequately (see Cook, Gelman and Rubin, \citeyear{CoGeRu2006}, for
elaboration).

While the link between $(\mathit{SN}, \mathit{SP})$ and $( \theta_{0},\theta_{1} )$
is exploited to advantage under a Bayesian analysis, it is problematic
for the FCI based on taking $\mathit{SN}$ and $\mathit{SP}$ as fixed values less than one.
In particular, the FCI is not defined for data sets with one or both
$\hat{\theta} _{i}$ falling outside the interval $(1-\mathit{SP},
\mathit{SN})$. Moreover, this can happen via sampling variation even if
the postulated values of $(\mathit{SN}, \mathit{SP})$ happen to be correct. Tu,
Litvak and Pagano (\citeyear{TuLiPa1994}, \citeyear{TuLiPa1995}) discussed this problem, and offered some
mitigating strategies when exposure prevalence (say, in a single
population) is the inferential target of interest. Such strategies yield
interval estimates for prevalence with an endpoint at zero or one, which
limits their utility for odds-ratio inference. In our case, the results
for estimator (ii) in Table \ref{tab5} are based on only the 81\% of sampled
parameter-data ensembles not giving rise to the aforementioned problem.
Perversely, this method is failing in situations where the data are most
suggestive that the guessed values of $(\mathit{SN}, \mathit{SP})$ might be wrong.
Put another way, the FCI fails on data sets where Bayesian intervals may
do particularly well via more prior-to-posterior updating of
$(\mathit{SN}, \mathit{SP})$.

As a final point concerning this example, we recognize that it is quite
reasonable to also study the frequentist properties of the Bayesian
interval estimator. This can become quite computationally burdensome,
however, if evaluation of frequentist coverage at many points in the
parameter space is desired: each point necessitates simulation of many
data sets, and each data set may require many MCMC iterations in order
to compute the interval estimate. Rather than pursuing this course, we
note that the simulation of parameter-data ensembles as used to evaluate
labwise coverage also yields information\break about frequentist coverage.

Thus, say that the frequentist coverage for parameter vector
$\theta^{*}$ is of interest. If $m$ parameter-data ensembles are
simulated, then we might consider the proportion $\alpha$ of ensembles
for which $\theta$ is closest to $\theta^{*}$ in some sense. Then the
empirical coverage for this subset of ensembles approximates the
frequentist coverage at $\theta^{*}$, with the approximation improving
as $\alpha \rightarrow 0$ and $m\alpha \rightarrow \infty$. Admittedly,
it may be computationally prohibitive to make the approximation error
very small, so we refer to the reported coverage as ``near-frequentist
coverage'' around $\theta^{*}$. Notwithstanding its approximate nature,
this can still reveal trends in frequentist coverage across the
parameter space.

To apply this to the present example, we extend the simulation size to
$m=100{,}000$ parameter-data ensembles, and set $\alpha =0.01$. Various
points $\theta^{*}$ in the parameter space are considered, by fixing
$r_{0}=0.10$, $r_{1}=0.15$, and then setting $\mathit{SN}$ and $\mathit{SP}$
at values corresponding to specific prior quantiles. Thus, we
investigate how the frequentist coverage depends on the compatibility
between the prior and the true $\mathit{SN}$ and $\mathit{SP}$ values. Results
appear in Table \ref{tab6}. We see under-coverage for $\mathit{SP}$ values which
are low in relation to the prior, and over-coverage when $\mathit{SP}$ or
$\mathit{SN}$ is high in relation to the prior. Generally, however, the
variation in frequentist coverage as $\mathit{SN}$ and $\mathit{SP}$ values
move around the region supported by the prior distribution seems quite
modest.

\section{Recommendations}\label{sec5}

The~above arguments and illustrations are intended to summarize and
explain in simple form several practical recommendations that we and
others have reached in the course of numerous theoretical studies,
simulations and real applications. Like others before us, we first
recommend forming prior distributions and then reporting Bayesian
interval estimates for parameters of interest, particularly in
nonidentified model contexts. Based on our investigations, however, we
further suggest that a special form of sensitivity analysis be carried
out as well.

Sensitivity analysis is conducted in much applied work; typically this
involves reporting multiple inferences corresponding to multiple models
and (for Bayesians) multiple prior distributions. While these analyses
are often better than standard reports of results from just one model,
the resulting collection of interval estimates leads to problems of
summarization and interpretation of the collection. Thus, we recommend
instead that one start with a single, relatively inclusive ``covering''
prior distribution that subsumes the diversity of opinions and
possibilities for the parameters. Then, as a safeguard, we would
evaluate the labwise coverage of Bayesian intervals arising from this
prior, for a variety of PGDs differing somewhat from the prior. If the
coverage does not fall much below nominal as the PGD deviates from the
prior, then we may argue that our statistical procedure is probably (in
the subjective judgmental sense) at least roughly calibrated, in the
across-study sense of labwise coverage. Otherwise, we may consider
ourselves alerted to a potentially serious miscalibration.

Table \ref{tab3}, in the context of prevalence surveys with nonresponse, provides
one example of studying the sensitivity of labwise coverage as the PGD
deviates from the prior distribution. We close with a further example
from a specific and well-developed scientific context.

\subsection{Example: Silica Exposure and Lung Cancer}

We revisit the investigation of Steenland and Greenland (\citeyear{StGr2004}) on the
relationship of silica exposure to lung cancer. In a cohort of 4626
industrial sand workers with high silica exposure, 109 lung-cancer
deaths were observed, compared to an expected count of 68.1 under the
null hypothesis of no association between silica exposure and lung
cancer. This comparison of the cohort to US population data is
adjusted for age, race, calendar time and sex. It is not adjusted for
smoking status though, because smoking histories were not collected for
this cohort.

Steenland and Greenland used prior information derived from other
studies in order to remedy this situation using both Monte Carlo
sensitivity analysis (MCSA) and Bayesian analysis. To describe this
analysis, let $\beta_{1}$ be the log relative risk of lung-cancer death
for silica exposure versus no exposure, within strata defined by smoking
behavior, and let $\beta_{2}$ and $\beta_{3}$ be log relative risks for
current smokers compared to never smokers, and former smokers compared
to never smokers. Assuming a log-linear model without products between
silica exposure and smoking effects, the observed death count can be
regarded as a Poisson realization with log-mean $\lambda$, where
\begin{eqnarray*}
\lambda &=& c + \beta_{1} + \log(p_{1} + p_{2}e^{\beta _{2}} +
p_{3}e^{\beta {}_{3}})\\
&&{} - \log( q_{1} + q_{2}e^{\beta _{2}} +
q_{3}e^{\beta _{3}} ).
\end{eqnarray*}
Here $c$ is a known offset obtained from population data ($c=\log 68.1$
in the present example), while $(p_{1}, p_{2}, p_{3})$ and
$(q_{1},q_{2},q_{3})$ are probability distributions over (never,
current, former) smokers, in the exposed and unexposed populations
respectively.\break This is a highly nonidentified model, with nine unknown
parameters involved in the mean function. Identification of the target
parameter $\beta_{1}$ can only be obtained via a strong assumption, for
example, that smoking behavior and occupational silica exposure are
unassociated, that is, $(p_{1}, p_{2}, p_{3}) = (q_{1},q_{2},q_{3})$,
which is known to be false. Thus, a~far more principled analysis
combines the Poisson model for data along with prior distributions for
$(\beta_{1},\beta_{2},\beta_{3}), (p_{1},p_{2},\break p_{3})$ and
$(q_{1},q_{2},q_{3})$.

Based on data from a large cohort study of smoking and lung cancer,
Steenland and Greenland took $\beta_{2} \sim N(\log(23.6), 0.094^{2})$
and independently $\beta_{3} \sim N(\log(8.7), 0.094^{2})$. They used
smoking data on a small sample of 199 workers to inform the prior $p
\sim \operatorname{Dirichlet}( 199\times(0.26, 0.40, 0.34))$, and used a large
national survey to inform the prior on $q$. This survey involved 56,000
subjects, but to account for various uncertainties it was discounted by
a factor of four to yield the prior $q \sim \operatorname{Dirichlet}( 14{,}000
\times(0.34, 0.35,\break 0.31))$. Steenland and Greenland used a very diffuse
prior on~$\beta_{1}$. This is not appropriate for investigating labwise
coverage, however, as some data sets simulated from parameters generated
under this prior will have implausibly low (i.e., zero) death counts,
while others will have implausibly large counts. Thus, for present
purposes we take the prior $\beta_{1} \sim\break N[0, \{ln(5)/2\}^{2}]$,
which puts most of its weight on relative risks between $1/5$ and 5. This
completes specification of the prior distribution.

Bayesian computation for the present situation is readily implemented in
a two-stage manner. First, an approximate posterior sample is simulated
by drawing $\lambda$ values ``as if'' $\lambda$ had a flat prior, and
independently drawing $(\beta_{2}, \beta_{3}, p_{1}, p_{2}, p_{3},
q_{1}, q_{2}, q_{3})$ values from their prior distribution. Second, this
posterior sample is ``made exact'' via importance sampling, which
recognizes the actual prior distribution in the $(\lambda, \beta_{2},
\beta_{3}, p_{1}, p_{2}, p_{3}, q_{1}, q_{2}, q_{3})$ parameterization.
Note that omitting the second step corresponds to the MCSA in Steenland
and Greenland. In the present example this second step has negligible
impact, though in general importance sampling can be used to convert
MCSA inferences to fully Bayesian inferences in situations where the two
do not agree so closely.

Applied to the cohort data, a 95\% equal-tailed BPCI for
$\exp(\beta_{1})$ is $(1.12, 1.73)$, which is very similar to the interval
reported by Steenland and Greenland using their slightly different
prior. For comparison, the analysis which ignores the confounding effect
of smoking gives the interval $(1.31, 1.91)$. This result is based on the
same prior for $\beta_{1}$ as above, with the presumption that $(p_{1},
p_{2}, p_{3}) = (q_{1}, q_{2}, q_{3})$. Thus, the impact of
acknowledging smoking as a confounder is to push the interval estimate
for $\beta_{1}$ toward (but not across) the null, and to widen the
interval by about 15\%. This widening is somewhat modest, since there is
relatively good prior data about smoking effects and smoking behavior in
the two populations and the association of smoking with silica exposure
in these data appears to be small.

We know that BPCIs based on this prior will have correct labwise
coverage for a PGD equal to the prior. We wish to see how far the
coverage deviates from nominal as the PGD deviates from the prior. We
thus examine eight PGDs, starting with the prior and considering all
possible combinations of:

\begin{longlist}[(iii)]
\item[(i)] shifting the prior mean for $\beta_{2}$ left or right by one
prior standard deviation;

\item[(ii)] shifting the prior mean for $\beta_{3}$ left or right by one
prior standard deviation;

\item[(iii)] discounting the prior on $(p_{1},p_{2},p_{3})$ by a
factor of two or (further) discounting the prior on
$(q_{1},q_{2},\break q_{3})$ by a factor of two.
\end{longlist}

Table \ref{tab7} gives coverage results using 95\% equal-tailed BPIs for
$\beta_{1}$. When the PGD equals the prior, the labwise coverage is
within simulation error of nominal, as theory dictates. As the model is
highly nonidentified, we are not surprised to see lower than nominal
coverage for most of the PGDs considered. We are pleasantly surprised,
however, to see that the loss of coverage is very mild. This adds
credence to the Bayesian results given by Steenland and Greenland
(\citeyear{StGr2004}).

\begin{table}
\caption{Labwise coverage of 95\% Bayesian intervals for $\beta_{1}$ as
the PGD varies, in the silica and lung cancer example. The~first row
gives coverage when the PGD equals the prior. The~remaining eight rows
give coverage when the PGD is an alteration of the prior. The~three-character code describes the alteration. The~first character ($+$ or
$-$) indicates whether the mean of $\beta_{2}$ is increased or
decreased, the second character does the same for the mean of~$\beta_3$, and the third character ($p$ or $q$) indicates whether the
prior on p or the prior on q is discounted. Results are based on 100,000
realizations, giving simulation error for coverage less than 0.1\%}
\label{tab7}
\begin{tabular*}{\columnwidth}{@{\extracolsep{\fill}}lc@{}}
\hline
\textbf{PGD} & \textbf{Coverage \%}\\
\hline
Prior & 94.8\\
$- - p$ & 92.6\\
$- - q$ & 94.8\\
$- + p$ & 93.1\\
$- + q$ & 95.2\\
$+ - p$ & 92.0\\
$+ - q$ & 94.5\\
$+ + p$ & 92.4\\
$+ + q$ & 94.8\\
\hline
\end{tabular*}
\end{table}

Based on examples as well as theoretical and simulation studies, we
recommend that PGD sensitivity analysis be used when inference based on
nonidentified models is required. No important sensitivity was seen in
the preceding example. Nonetheless, high sensitivity to plausible PGD
specifications would have suggested that the full model (including those
for the prior distribution and data-generating mechanism) had
inadequately captured posterior uncertainty given the actual prior
uncertainty of the analysts, and that interval estimates from the model
could be seriously miscalibrated. Hence, as with failed regression
diagnostics, we would find ourselves advised to revise our model rather
than rely on it.

Of course, this advice raises classic issues of the impact of post-data
model revision based on diagnostics, long recognized as a challenge for
applied Bayesians as well as for applied frequentists (Box, \citeyear{Bo1980}). We
thus regard these issues as an important direction for further research
in our proposed approach.

\section*{Acknowledgments}

Paul Gustafson's research was supported by grants from the Natural
Sciences and Engineering Research Council of Canada and the Canadian
Institutes for Health Research (Funding Reference Number 62863).

\vspace*{-2pt}

\end{document}